# Phase Slips and the Eckhaus Instability

*J.-P. Eckmann*[1,2], *Th. Gallay*[3], *and C.E. Wayne*[4]

[1] Dépt. de Physique Théorique, Université de Genève, CH-1211 Genève 4, Switzerland
[2] Section de Mathématiques, Université de Genève, CH-1211 Genève 4, Switzerland
[3] Dépt de Mathématiques, Université de Paris-Sud, bât 425, F-91405 Orsay, France
[4] Dept. of Mathematics, Penn. State University, State College PA 16802, USA

**Abstract.** We consider the Ginzburg-Landau equation, $\partial_t u = \partial_x^2 u + u - u|u|^2$, with complex amplitude $u(x,t)$. We first analyze the phenomenon of phase slips as a consequence of the *local* shape of $u$. We next prove a *global* theorem about evolution from an Eckhaus unstable state, all the way to the limiting stable finite state, for periodic perturbations of Eckhaus unstable periodic initial data. Equipped with these results, we proceed to prove the corresponding phenomena for the fourth order Swift-Hohenberg equation, of which the Ginzburg-Landau equation is the amplitude approximation. This sheds light on how one should deal with local and global aspects of phase slips for this and many other similar systems.

## 1. Introduction

In the last few years we have seen continuous progress in the rigorous understanding of pattern-forming systems, describing successfully fronts, modulation equations, nucleation mechanisms and scaling phenomena. The aim of the present paper is to add to this body of results a rigorous study of "phase slips," that is, of solutions whose number of zeros varies as a function of time. The problem of the variation of the number of zeros in parabolic PDE's has been studied earlier, see, e.g. [A], and the references therein. Our paper will deal with two novel aspects of this problem, one local and the other global, which use complex order parameters and apply also to higher order differential operators.

The *local theory* is based on the prototype of the Ginzburg-Landau (GL) equation:

$$\partial_t u(x,t) = \partial_t^2 u(x,t) + u(x,t) - u(x,t)|u(x,t)|^2 , \tag{1.1}$$

where $u : \mathbf{R} \times \mathbf{R}^+ \to \mathbf{C}$. It allows one to predict from the local shape of the solution at time $t = 0$ whether there will be a phase-slip—a 0 of $|u(\cdot,t)|$—at some later time $t$. The condition is that

$$u(x,0) \approx \tfrac{1}{2}ax^2 + ibx - ad , \tag{1.2}$$

where $a > 0$, and $\operatorname{Re} b \neq 0$. We can then show the appearance of a phase slip for sufficiently small $d = d(a,b) > 0$. (This condition says that the initial condition is "just before" a phase slip.) The point that we wish to emphasize is that while we have stated this result as a theorem about the initial value problem, in fact it implies that whenever any solution of the GL equation gets close to zero it will undergo a phase slip by the mechanism we describe, provided that it has a non-zero imaginary part, and that its curvature is small and has a prescribed sign. We conjecture that except for very special (i.e., "non-generic," in the appropriate topology) solutions, all phase slips occur via the mechanism we describe. To see what we mean by "special," note that real valued solutions, for instance, do not satisfy our criteria, but from the point of view of considering the GL equation as an amplitude equation, which is the source of most of our interest, real solutions are very atypical. The proof of the above result is given in Theorem 2.7. It is a perturbative argument, and therefore it can be extended to other, more general problems, of which the GL equation is an "amplitude equation." We illustrate this in Section 4 for the specific example of the Swift-Hohenberg (SH) equation

$$\partial_t U = \left(3\varepsilon^2 - (1 + \partial_x^2)^2\right)U - U^3 . \tag{1.3}$$

We consider initial data of the form

$$U(x,0) = 2\varepsilon \operatorname{Re}\left(u(\tfrac{\sqrt{3}}{2}\varepsilon x, 0)e^{ix}\right) , \tag{1.4}$$

where $u$ satisfies the Ginzburg-Landau slip condition Eq.(1.2). Then, if $\varepsilon > 0$ is sufficiently small, the solution $U(\cdot,t)$ will have a phase slip, i.e., a pair of roots of $U$ will appear or disappear at some time $t \approx d/(3\varepsilon^2)$. This is the way in which the phase slip of the amplitude part of an equation translates to the equation itself, see Theorem 4.1.

In many experiments phase slips are produced not by putting the system near the state described by Eq.(1.2) but rather by starting with the system in an unstable state, perturbing it slightly and waiting for it to undergo a phase slip. This is a *global* problem, and, obviously more care and control are needed for a rigorous treatment than in the local problem. We study this global aspect in the setting of the celebrated Eckhaus instability for the GL equation in Theorem 3.2. In order to have good control, we consider an Eckhaus unstable stationary solution $u(x,0) = \sqrt{1-q_0^2}\, e^{iq_0 x}$ which we perturb by adding a *single frequency* $\varepsilon e^{iq_1 x}$, or, more generally, an initial condition of the form

$$\sqrt{1-q_0^2}\, e^{iq_0 x} + \sum_{n \in \mathbf{Z}} \varepsilon_n e^{i(q_0 + n(q_1 - q_0))x} \ , \tag{1.5}$$

with $|\varepsilon_n|$ small. (Note that the subspace of such functions is left invariant by the GL evolution Eq.(1.1).) If $q_0$ and $q_1$ are suitably chosen as described in Section 3, we can show that the time-evolution must bring the solution from the initial state to the state $\sqrt{1-q_1^2}\, e^{iq_1 x}$ (up to a phase). Thus, this demonstrates the full evolution of an Eckhaus unstable state. In the course of this evolution, the wavelength will change from $2\pi/q_0$ to $2\pi/q_1$ and phase slips will occur.

In Section 4, we sketch, in quite some detail, the proof of the following novel *global result*: Consider the SH equation with initial data

$$U(x,0) = 2\varepsilon \mathrm{Re}\left(A(\tfrac{\sqrt{3}}{2}\varepsilon x, 0)e^{ix}\right) \ . \tag{1.6}$$

Assume the initial data Eq.(1.6) have an amplitude $A$ which is Eckhaus unstable and is perturbed as in Eq.(1.5). Then the SH equation will *globally* change its amplitude modulation as time goes on, and come close to a new periodic state whose amplitude has wavevector $q_1$. These results are based on recent estimates of the relation between the evolution of the SH equation and the GL equation [CE2, Sch].

## 2. The Ginzburg-Landau Equation

In this section we consider the Ginzburg-Landau equation,

$$\partial_t u(x,t) = \partial_x^2 u(x,t) + u(x,t) - u(x,t)|u(x,t)|^2 \ , \tag{2.1}$$

where $u : \mathbf{R} \times \mathbf{R}^+ \to \mathbf{C}$. Our understanding of the phase-slip mechanism is based on the following simple observation. Assume that $u(x,t)$ is a solution satisfying $u(0,t_0) = 0$ for some $t_0 > 0$. Then, using the well-known regularity properties of solutions of parabolic equations, we can expand $u$ near $x = 0, t = t_0$ as

$$u(x, \tau + t_0) = \tfrac{1}{2}\alpha x^2 + \beta x + \gamma \tau + \mathcal{O}(|\tau|^{3/2} + |x|^3) \ ,$$

where $\alpha, \beta, \gamma \in \mathbf{C}$ and the term of the form $\tau x$ has been absorbed in the remainder. Since $u$ satisfies Eq.(2.1) and vanishes at $x = 0, t = t_0$, we deduce that $\alpha = \gamma$. Also, by the phase invariance of the equation, we may assume without loss of generality that $\alpha \in \mathbf{R}$, $\alpha \geq 0$. Therefore, we expect the function

$$u_s(x,t) = \tfrac{1}{2}\alpha x^2 + \beta x + \alpha(t - t_0), \quad \alpha \in \mathbf{R}^+, \quad \beta \in \mathbf{C},$$

to be a good approximation to $u(x,t)$, near $x = 0, t = t_0$. Note that $u_s(x,t)$ is an exact solution of the linear equation $\partial_t u(x,t) = \partial_x^2 u(x,t)$. Furthermore, if $\alpha > 0$ and $\operatorname{Im}\beta \neq 0$ (which is generically expected to hold), then $u_s(x,t) = 0$ if and only if $x = 0$ and $t = t_0$. A sequence of plots indicating the phase slip for $u_s$ is given in Fig. 1.

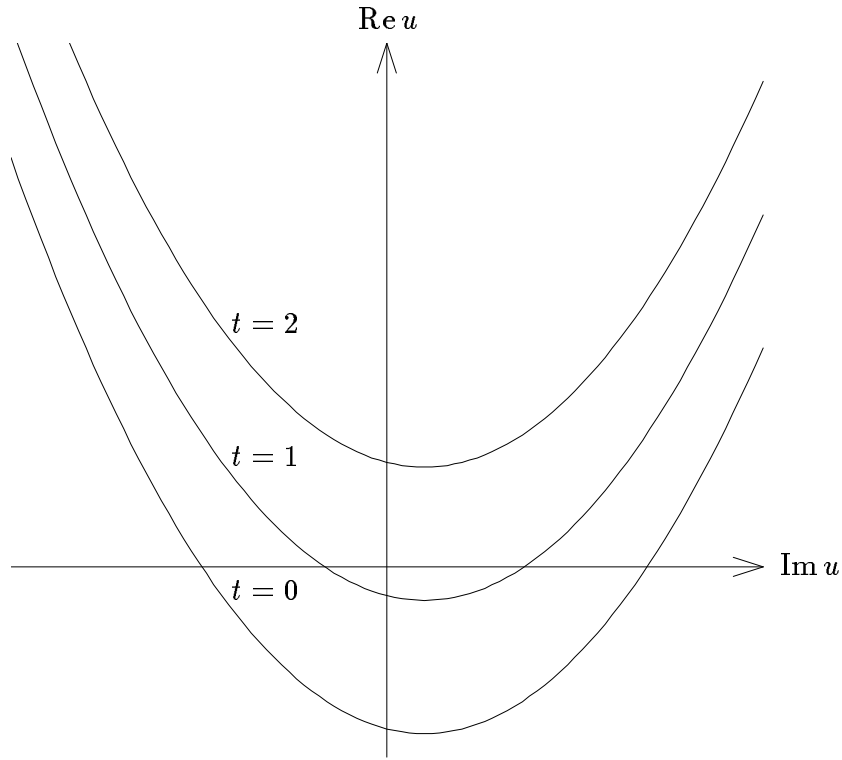

**Fig. 1**: The real and imaginary parts of the function $u_s(x,t)$ for 3 successive times.

We now change our perspective and formulate the problem in terms of initial data. Let

$$u_0(x) = \tfrac{1}{2}ax^2 + ibx - ad, \tag{2.2}$$

with $a > 0$, $d > 0$, $\operatorname{Re} b \neq 0$. We already know that the corresponding solution

$$u_0(x,t) = \tfrac{1}{2}ax^2 + ibx + a(t-d) \tag{2.3}$$

of the heat equation undergoes a phase slip at $x = 0$, $t = d$. The aim of this section is to show that this property remains true for any solution $u(x,t)$ of the GL equation with initial condition

$$u(x,0) = u_0(x) + \mathcal{O}(|x|^3), \tag{2.4}$$

near $x = 0$, provided $d$ is sufficiently small. Our main result can be outlined as follows:

**Proposition 2.1.** *If $d$ is small, and $u(x,0)$ is given by Eq.(2.4) near $x = 0$, then the solution $u(x,t)$ of the Ginzburg-Landau equation Eq.(2.1) will have a phase slip at time $t^* = d + \mathcal{O}(d^{3/2})$ and at $x^* = \mathcal{O}(d^{3/2})$.*

**Remark.** A precise formulation will be given in Theorem 2.7 below.

**Proof.** We shall give the proof for the slightly more general equation

$$\partial_t u(x,t) = \partial_x^2 u(x,t) + \mathcal{N}(u(x,t)), \qquad (2.5)$$

where $\mathcal{N}(z) = zP(|z|^2)$ and $P$ is a real polynomial whose leading coefficient is strictly negative. Under these assumptions, every bounded solution $u(x,t)$ of Eq.(2.5) satisfies for sufficiently large $t$:

$$\|u(\cdot,t)\|_{\mathcal{C}^3} \leq U^*, \qquad (2.6)$$

see [CE2]. In the sequel, we always assume that the bound (2.6) holds for all $t \geq 0$. Then there is a constant $N_1$ such that

$$|\mathcal{N}(z)| \leq N_1|z|, \quad |\partial_z^\ell \mathcal{N}(z)| \leq N_1, \quad \text{for } |z| \leq U^* \text{ and } \ell = 1, 2, 3. \qquad (2.7)$$

As announced in Eq.(2.4), we consider initial data of the form $u(x,0) = u_0(x) + w_0(x)$, with $u_0(x)$ of the form Eq.(2.2) (on the whole line) and $w_0(x) = \mathcal{O}(|x|^3)$ near $x = 0$. More precisely, we assume that there is a constant $W^* > 0$ such that

$$|w_0(x)| \leq W^*|x|^3, \quad |w_0'(x)| \leq W^*|x|^2, \quad |w_0''(x)| \leq W^*|x|, \quad x \in \mathbf{R}. \qquad (2.8)$$

Setting $\mathcal{L} = \partial_x^2$ and applying the Duhamel formula, we obtain the following representation for the solution:

$$\begin{aligned} u(x,t) &= \left(e^{t\mathcal{L}}(u_0 + w_0)\right)(x) + \int_0^t ds \left(e^{(t-s)\mathcal{L}}\mathcal{N}(u(\cdot,s))\right)(x) \\ &= u_0(x,t) + w(x,t) + z(x,t). \end{aligned} \qquad (2.9)$$

By construction, $u_0(x,t)$ vanishes at $t = d$ and $x = 0$ (and only there). Using the Implicit Function Theorem, we shall show that the full solution $u(x,t)$ has a unique zero inside the region

$$\mathcal{R}_d = \{(x,t) \mid x^2 \leq d, \ 0 \leq t \leq 2d\}, \qquad (2.10)$$

if $d$ is sufficiently small. We bound the three terms in Eq.(2.9) as follows:

**Bounds on $u_0$.** Since $u_0(x,t)$ is explicitly given by Eq.(2.3), it is straightforward to prove the following

**Lemma 2.2.** *There is a constant $K_0 = K_0(a,b)$ such that for $d \in [0,1]$ one has*

$$|u_0(x,t)| \leq K_0 d^{1/2}, \quad |\partial_x u_0(x,t)| \leq K_0, \quad |\partial_t u_0(x,t)| \leq K_0, \qquad (2.11)$$

for all $(x,t) \in \mathcal{R}_d$.

**Bounds on $w$.** Here, we exploit the fact that $w_0$ vanishes at the origin. For any function $f$, we have the explicit representation

$$(e^{t\mathcal{L}}f)(x) = (4\pi t)^{-1/2} \int_{-\infty}^{\infty} dy\, e^{-(x-y)^2/(4t)} f(y) \,. \tag{2.12}$$

If $|f(x)|$ is bounded by $|x|^m$, $m \geq 0$, we find

$$\begin{aligned}
\left|(e^{t\mathcal{L}}f)(x)\right| &\leq C t^{-1/2} \int_{-\infty}^{\infty} dy\, e^{-(x-y)^2/(4t)} |y|^m \\
&\leq C \int_{-\infty}^{\infty} d\xi\, e^{-\xi^2/4} |x + \xi t^{1/2}|^m \\
&\leq C \left||x| + t^{1/2}\right|^m \,.
\end{aligned} \tag{2.13}$$

Here and in the sequel, $C$ denotes a constant which can vary from equation to equation. In view of (2.8), we can apply Eq.(2.13) to $w_0$ and its derivatives and obtain

$$|w(x,t)| \leq C \left||x| + t^{1/2}\right|^3 \,,\quad |\partial_x w(x,t)| \leq C \left||x| + t^{1/2}\right|^2 \,,\quad |\partial_x^2 w(x,t)| \leq C \left||x| + t^{1/2}\right| \,.$$

Therefore we have shown:

**Lemma 2.3.** *There is a constant $K_1 = K_1(W^*)$ such that for $d \in [0,1]$ one has*

$$|w(x,t)| \leq K_1 d^{3/2} \,,\quad |\partial_x w(x,t)| \leq K_1 d \,,\quad |\partial_t w(x,t)| \leq K_1 d^{1/2} \,, \tag{2.14}$$

*for all $(x,t) \in \mathcal{R}_d$.*

**Bounds on $z$.** To bound the nonlinear term, we use the integral equation

$$z(x,t) = \int_0^t ds\, \left(e^{(t-s)\mathcal{L}} \mathcal{N}(u_0(\cdot,s) + w(\cdot,s) + z(\cdot,s))\right)(x) \,, \tag{2.15}$$

which follows immediately from Eq.(2.9). Note that the right-hand side depends on the solution $u(x,t)$ on the whole real line, whereas we are only interested in bounding $z$ for $(x,t) \in \mathcal{R}_d$. We use the following "localization" lemma:

**Lemma 2.4.** *There is a constant $K_2$ such that for all $f \in \mathcal{C}^1$ and all $t > 0$ one has*

$$|(e^{t\mathcal{L}}f)(x) - f(x)| \leq K_2 \sqrt{t} \|f\|_{\mathcal{C}^1} \,.$$

**Remark.** To be specific, we define

$$\|f\|_{\mathcal{C}^1} = \max\left(\sup_x |f(x)|\,,\ \sup_x |f'(x)|\right) \,.$$

**Proof.** Using the representation (2.12) and the Mean Value Theorem, we obtain

$$\begin{aligned}
\left(e^{t\mathcal{L}}f\right)(x) &= \frac{1}{(4\pi t)^{1/2}} \int_{-\infty}^{\infty} d\xi \, e^{-\xi^2/(4t)} f(x+\xi) \\
&= f(x) + \frac{1}{(4\pi t)^{1/2}} \int_{-\infty}^{\infty} d\xi \, e^{-\xi^2/(4t)} \left(f(x+\xi) - f(x)\right) \\
&= f(x) + \frac{1}{(4\pi t)^{1/2}} \int_{-\infty}^{\infty} d\xi \, e^{-\xi^2/(4t)} \xi f'(\vartheta(x,\xi)) \,,
\end{aligned}$$

where $\vartheta(x,\xi)$ is some point between $x$ and $x+\xi$. The assertion follows.

We apply this lemma to bound the right-hand side of Eq.(2.15). Using Eqs.(2.6), (2.7), we find

$$\begin{aligned}
\left|\left(e^{(t-s)\mathcal{L}}\mathcal{N}(u(\cdot,s))\right)(x) - \mathcal{N}(u(x,s))\right| &\leq K_2\sqrt{t-s}\,\|\mathcal{N}(u(\cdot,s))\|_{C^1} \\
&\leq K_2\sqrt{t-s}\,N_1\|u(\cdot,s)\|_{C^1} \leq K_2\sqrt{t-s}\,N_1 U^*.
\end{aligned}$$

Integrating this bound over $s$, we obtain

$$\left|\int_0^t ds\,\left(e^{(t-s)\mathcal{L}}\mathcal{N}(u(\cdot,s))\right)(x) - \int_0^t ds\,\mathcal{N}(u(x,s))\right| \leq K_3 t^{3/2} \,, \tag{2.16}$$

where $K_3 = (2/3)K_2 N_1 U^*$.

On the other hand, using Lemma 2.2, Lemma 2.3 and Eqs.(2.7), (2.9), we have for all $(x,s) \in \mathcal{R}_d$

$$\begin{aligned}
|\mathcal{N}(u(x,s))| &\leq N_1(|u_0(x,s)| + |w(x,s)| + |z(x,s)|) \\
&\leq N_1\left(K_0 d^{1/2} + K_1 d^{3/2} + |z(x,s)|\right) \,,
\end{aligned}$$

and hence

$$\int_0^t ds\,|\mathcal{N}(u(x,s))| \leq N_1(K_0 + K_1)d^{1/2}t + N_1 \int_0^t ds\,|z(x,s)| \,. \tag{2.17}$$

Finally, combining Eqs.(2.15)–(2.17), we see that

$$|z(x,t)| \leq K_4 d^{1/2} t + N_1 \int_0^t ds\,|z(x,s)| \,,$$

for all $(x,t) \in \mathcal{R}_d$, where $K_4 = \sqrt{2}K_3 + N_1(K_0 + K_1)$. Therefore, it follows from Gronwall's inequality that

$$|z(x,t)| \leq K_4 d^{1/2} t \exp(N_1 t) \leq 2K_4 d^{3/2} \exp(2N_1 d) \,,$$

for all $(x,t) \in \mathcal{R}_d$.

To bound the derivatives of $z$, we differentiate Eq.(2.15) with respect to $x$ and obtain integral equations for $\partial_x z$ and $\partial_x^2 z$. Repeating exactly the same estimates, we find that $|\partial_x z(x,t)| \leq Cd$ and $|\partial_x^2 z(x,t)| \leq Cd^{1/2}$ for all $(x,t) \in \mathcal{R}_d$. Since $\partial_t z(x,t) = \partial_x^2 z(x,t) + \mathcal{N}(u(x,t))$, we see that $|\partial_t z(x,t)| \leq Cd^{1/2}$ when $(x,t) \in \mathcal{R}_d$. Thus, we have shown:

**Lemma 2.5.** *There is a constant $K_5 = K_5(a, b, U^*, W^*)$ such that for $d \in [0,1]$ one has*

$$|z(x,t)| \leq K_5 d^{3/2}, \quad |\partial_x z(x,t)| \leq K_5 d, \quad |\partial_t z(x,t)| \leq K_5 d^{1/2}, \tag{2.18}$$

*for all $(x,t) \in \mathcal{R}_d$.*

We summarize Lemma 2.3 and Lemma 2.5 in the following

**Proposition 2.6.** *Assume that $u(x,0) = u_0(x) + w_0(x)$, with $u_0$ given by Eq.(2.2) and $w_0$ satisfying Eq.(2.8). Assume that the solution $u(x,t)$ of Eq.(2.1) satisfies $\|u(\cdot,t)\|_{\mathcal{C}^3} \leq U^*$ for all $t \geq 0$. Then there is a constant $K^* = K^*(a, b, U^*, W^*)$ so that, for all $d \in [0,1]$, the inequalities*

$$|u(x,t) - (\tfrac{1}{2}ax^2 + ibx + a(t-d))| \leq K^* d^{3/2},$$
$$|\partial_x u(x,t) - (ax + ib)| \leq K^* d,$$
$$|\partial_t u(x,t) - a| \leq K^* d^{1/2},$$

*hold on $\mathcal{R}_d$.*

**Existence of the phase slip.** In view of Proposition 2.6, the existence of a phase slip is now a straightforward application of the Implicit Function Theorem. Indeed, we can view the function $u(x,t) = (\operatorname{Re} u(x,t), \operatorname{Im} u(x,t))$ as a (smooth) map from $\mathcal{R}_d$ to $\mathbf{R}^2$. By Proposition 2.6, we know that $u(0,d) = \mathcal{O}(d^{3/2})$ and the derivative $Du(x,t)$ satisfies

$$Du(x,t) = \begin{pmatrix} ax - \operatorname{Im} b & a \\ \operatorname{Re} b & 0 \end{pmatrix} + \mathcal{O}(d^{1/2}).$$

Therefore, since $a > 0$ and $\operatorname{Re} b \neq 0$, $Du(x,t)$ is close to a non-singular matrix if $d$ is sufficiently small. This implies that $u$ is one-to-one from $\mathcal{R}_d$ onto a neighborhood of size $\mathcal{O}(d)$ around $u(0,d)$ in $\mathbf{R}^2$. In particular, there is a unique point $(x^*, t^*) \in \mathcal{R}_d$ such that $u(x^*, t^*) = 0$. Applying the implicit function theorem, we find

$$\begin{pmatrix} x^* \\ t^* \end{pmatrix} = \begin{pmatrix} 0 \\ d \end{pmatrix} - (Du(0,d))^{-1} \begin{pmatrix} \operatorname{Re} u(0,d) \\ \operatorname{Im} u(0,d) \end{pmatrix} + \mathcal{O}(d^2).$$

Thus, our final result is:

**Theorem 2.7.** *Assume that $u(x,0) = u_0(x) + w_0(x)$, with $u_0$ given by Eq.(2.2) and $w_0$ satisfying Eq.(2.8). Assume that the solution $u(x,t)$ of the Ginzburg-Landau equation with initial data $u(x,0)$ satisfies $\|u(\cdot,t)\|_{\mathcal{C}^3} \leq U^*$ for all $t \geq 0$. Then there is a constant $d^* = d^*(a, b, U^*, W^*) > 0$ such that, for all $d \in [0, d^*]$, $u(x,t)$ has exactly one phase slip in the region $\mathcal{R}_{d^*}$. This phase slip will take place at time $t^* = d + \mathcal{O}(d^{3/2})$ and at $x^* = 0 + \mathcal{O}(d^{3/2})$.*

## 3. The Eckhaus Instability

The considerations of the preceding section were local in nature. Namely, we identified a local shape of the amplitude that looks "like just before a phase slip" and then showed that indeed a phase slip is going to occur in a well controlled region of space-time. Now, we wish to address the more difficult question of global aspects. We will give a mathematical *proof* of a phenomenon which is universally believed to hold in a very general context. However, our own results here apply in a more restricted setting, due to the requirement of rigor.

We consider again the Ginzburg-Landau equation, and more specifically perturbations of the space periodic time-independent solutions

$$s_{q_0}(x) = e^{iq_0 x}\sqrt{1-q_0^2}, \quad q_0 \in [-1,1].$$

As explained in the Introduction, we shall perturb $s_{q_0}(x)$ by adding essentially a single frequency $\varepsilon e^{iq_1 x}$, and we shall show that, under suitable assumptions on the parameters $q_0$ and $q_1$, this perturbation will grow and eventually drive the system to the new stationary solution $s_{q_1}(x)$. Since the wave-lengths $2\pi/q_0$, $2\pi/q_1$ are different, phase slips will necessarily occur during this transition.

We first recall that the spectrum of the linearized equation around $s_{q_0}(x)$, considered on the subspace of perturbations which are of the form

$$u(x) = e^{iq_0 x}\left(Ae^{i\delta x} + \bar{B}e^{-i\delta x}\right), \quad A,B \in \mathbf{C},$$

is given by

$$\lambda_\pm(q_0,\delta) = -(1-q_0^2) - \delta^2 \pm \sqrt{(1-q_0^2)^2 + 4q_0^2\delta^2}. \tag{3.1}$$

The quantity $\lambda_+$ can take positive values if and only if $q_0^2 > 1/3$, which is the celebrated instability region of Eckhaus [CE1]. Therefore, the perturbation $\varepsilon e^{iq_1 x}$ will grow at onset if we choose $q_0, q_1$ so that $q_1 = q_0 - \delta$ and $\lambda_+(q_0,\delta) > 0$. We thus take $q_0 > 1/\sqrt{3}$ and $0 < \delta < \sqrt{6q_0^2 - 2}$.

Of course, this linear condition does not guarantee that the system will converge to the stationary solution $s_{q_1}(x)$ as $t \to \infty$. One reason for this is that Fourier modes other than $e^{iq_0 x}, e^{iq_1 x}$ will be excited by the nonlinearity, some of them possibly in the Eckhaus stable region, and it is not clear a priori which one will eventually "win." To illustrate this point, let us fix $q_0 = \sqrt{2/5}$, $\delta = q_0/2$, and consider the solution $u(x,t)$ of the GL equation with initial condition $u(x,0) = s_{q_0}(x) + \varepsilon e^{iq_1 x}$ for some $\varepsilon \ll 1$. As is easily seen, $u(x,t)$ is of the form

$$u(x,t) = \sum_{n \in \mathbf{Z}} u_n(t)e^{iq_n x}, \quad q_n = q_0 - n\delta, \tag{3.2}$$

where $u_0(0) = \sqrt{1-q_0^2}$, $u_1(0) = \varepsilon$ and $u_n(0) = 0$ otherwise. Our choice of $q_0, \delta$ implies that $e^{iq_1 x}$ is the *most unstable* Fourier mode around $s_{q_0}(x)$ (i.e., $\lambda_+(q_0,\delta)$ is maximal for $\delta = q_0/2$), whereas $e^{iq_2 x}$ is linearly *not excited at all*. This is visible in Fig. 2.

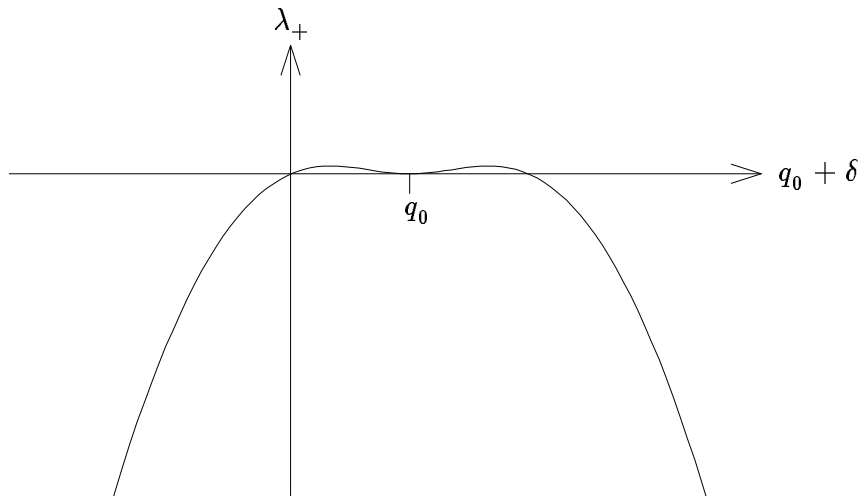

**Fig. 2**: The eigenvalue $\lambda_+(q_0, \delta)$, describing the instability of a perturbation with wave vector $q_0 + \delta$ of the periodic stationary solution $s_{q_0}$. The value $q_0 = \sqrt{2/5}$ is chosen such that $\lambda_+$ is maximal for $\delta = \pm q_0/2$. It is 0 for $\delta = \pm q_0$.

Nevertheless, a numerical calculation indicates that $u_2(t)$ acquires a significant size as time goes on, see Fig. 3.

Moreover, $u(x,t)$ does converge to $s_{q_1}(x)$ as $t \to \infty$, contrary to the rather common belief that the most unstable Fourier more around $u = 0$ (in our case, $e^{iq_2 x}$) will always win if it is excited. This example shows that the global evolution of the perturbation is not correctly described by the local analysis around the starting point, and that the final state is difficult to predict in general. However, under suitable assumptions on the parameters $q_0, q_1$, these difficulties can be overcome and the existence of a trajectory connecting $s_{q_0}$ to $s_{q_1}$ can be effectively proved, as we now proceed to show.

We begin by defining the function space in which we shall control the evolution of the system. According to Eq.(3.2), we let $\mathcal{U}_{q_0,\delta}$ be the Hilbert space of functions $u(x)$ of the form

$$u(x) = \sum_{n \in \mathbf{Z}} u_n e^{i(q_0 - n\delta)x}, \quad u_n \in \mathbf{C}, \tag{3.3}$$

equipped with the scalar product

$$(u,v)_\mathcal{U} = \frac{1}{T}\int_0^T (\bar{u}v + \bar{u}'v')\,dx = \sum_{n \in \mathbf{Z}} (1 + (q_0 - n\delta)^2)\,\bar{u}_n v_n,$$

where $T = 2\pi/\delta$. It is easy to see that $\mathcal{U}_{q_0,\delta}$ is invariant under the Ginzburg-Landau evolution

$$\dot{u} = u'' + u - u|u|^2, \tag{3.4}$$

and that $u \in \mathcal{U}_{q_0,\delta}$ if and only if $e^{-iq_0 x}u(x)$ is periodic of period $T$ and belongs to the Sobolev space $\mathrm{H}^1([0,T])$. So, $\mathcal{U}_{q_0,\delta}$ is isomorphic to the space

$$\mathrm{H}^1_{q_0}([0,T]) = \{u \in \mathrm{H}^1([0,T])\,|\,u(T) = e^{iq_0 T}u(0)\}.$$

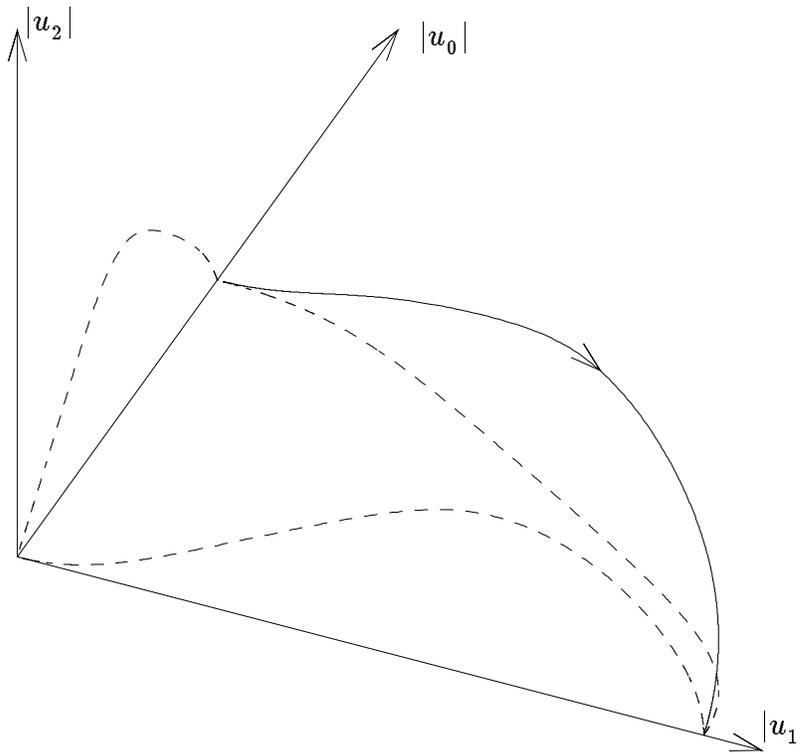

**Fig. 3**: Time evolution of the three main Fourier modes $u_n(t)$ of $u(x,t)$ in the case $q_0 = \sqrt{2/5}$, $\delta = q_0/2$. The periodic state $s_{q_0}(x)$ is slightly perturbed by $\varepsilon e^{iq_1 x}$, where $q_1 = q_0 - \delta$, and the system undergoes a transition to the state $s_{q_1}(x)$. In the intermediate time lapse, the amplitude $u_2$ acquires a significant size, about 10% of the total amplitude. The dashed lines are the three projections of the solid curve onto the coordinate planes.

Next, we note that the Ginzburg-Landau equation (3.4) is a "gradient system" in $\mathcal{U}_{q_0,\delta}$, in the sense that $\dot{u} = -\delta H/\delta u$, where

$$H(u) = \frac{1}{T}\int_0^T \left(\tfrac{1}{2}|u'|^2 - \tfrac{1}{2}|u|^2 + \tfrac{1}{4}|u|^4\right) dx, \quad T = \frac{2\pi}{\delta}.$$

In particular, the "energy" $H(u)$ is a decreasing function of time when $u$ is evolved with the GL equation. Our idea is to use this functional to show that any perturbation of $s_{q_0}(x)$ in $\mathcal{U}_{q_0,\delta}$ which decreases the energy $H$ will drive the system to the stationary solution $s_{q_1}(x)$. In particular, any initial condition $u(x,0)$ belonging to the unstable manifold of $s_{q_0}(x)$ in $\mathcal{U}_{q_0,\delta}$ converges to $s_{q_1}(x)$ as $t \to \infty$.

For our argument to work, it is essential that $s_{q_1}$ is *the only stationary solution $u$ of Eq.(3.4) in $\mathcal{U}_{q_0,\delta}$ satisfying $H(u) < H(s_{q_0})$*. Thus, in addition to $q_0 > 1/\sqrt{3}$ and $\delta < \sqrt{6q_0^2 - 2}$, we have to assume that $q_0 \leq \delta$, for if $q_0 > \delta$, then the wave-number $q_2 = q_0 - 2\delta$ verifies $|q_2| < q_0$, so that the stationary solution $s_{q_2}$ (which belongs to $\mathcal{U}_{q_0,\delta}$) satisfies $H(s_{q_2}) < H(s_{q_0})$. Note, however, that the assumption $q_0 \leq \delta$ excludes some interesting cases, like the one represented in Fig. 3.

According to these remarks, the optimal parameter range within the scope of our method is

$$\mathcal{E} = \{(q_0, \delta) \mid q_0 \leq \delta < \sqrt{6q_0^2 - 2} \text{ and } \sqrt{2/5} < q_0 \leq 1\}.$$

This domain is represented in Fig. 4 in the $q_0, q_1$ plane, where $q_1 = q_0 - \delta$. Although our results probably hold for all $(q_0, \delta) \in \mathcal{E}$, they are somewhat easier to prove when $\delta > 1$, and for simplicity we shall restrict ourselves to this case. Thus, we define

$$\hat{\mathcal{E}} = \{(q_0, \delta) \mid 1 < \delta < \sqrt{6q_0^2 - 2} \text{ and } 1/\sqrt{2} < q_0 < 1\} \subset \mathcal{E}.$$

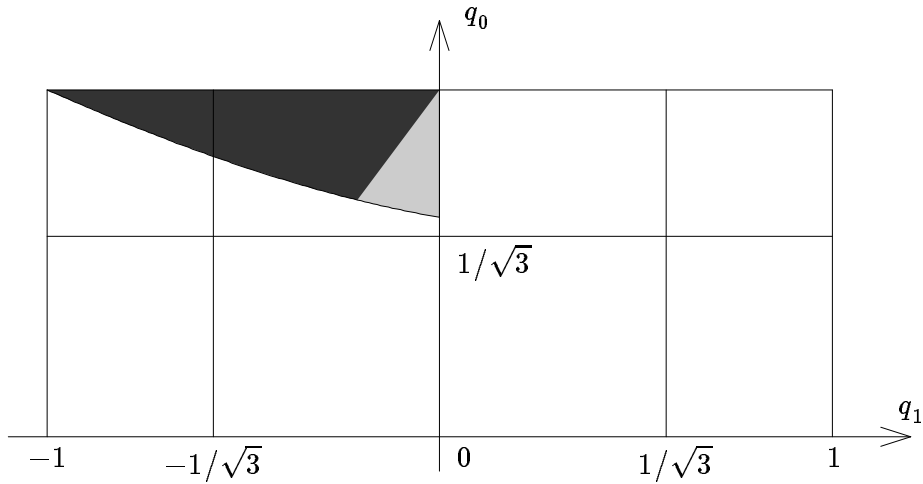

**Fig. 4**: The regions of the $q_1, q_0$ plane corresponding to the domains $\mathcal{E}$ (shaded) and $\hat{\mathcal{E}} \subset \mathcal{E}$ (black). For all $q_0, q_1$ in the black region, we prove the existence of an Eckhaus transition from $s_{q_0}$ to $s_{q_1}$, and we believe that the same is true for the whole shaded region.

**Proposition 3.1.** *If $(q_0, \delta) \in \hat{\mathcal{E}}$, then the only elements $u$ in $\mathcal{U}_{q_0,\delta}$ which are stationary solutions of the Ginzburg-Landau equation and whose energy $H(u)$ satisfies $H(u) < H(s_{q_0})$ are the functions*

$$u(x) = e^{i\varphi}\sqrt{1 - q_1^2}\, e^{iq_1 x},$$

*where $q_1 = q_0 - \delta$ and $\varphi \in [0, 2\pi)$.*

Using this result (which will be proved below), we now state and prove the main result of this section:

**Theorem 3.2.** *Let $(q_0, \delta) \in \hat{\mathcal{E}}$. Let $u^{(0)} \in \mathcal{U}_{q_0,\delta}$, and assume that $H(u^{(0)}) < H(s_{q_0})$. Then, as $t \to \infty$ the solution of the GL equation with initial data $u^{(0)}$ converges to*

$$u(x) = e^{i\varphi}\sqrt{1 - q_1^2}\, e^{iq_1 x} \qquad \text{for some } \varphi \in [0, 2\pi),$$

where $q_1 = q_0 - \delta$.

**Remark.** Although it is linearly unstable, the fixed point $s_{q_0}$ has an infinite-dimensional center-stable manifold $\mathcal{V}_{cs}$ in $\mathcal{U}_{q_0,\delta}$, so there exist many perturbations of $s_{q_0}$ in $\mathcal{U}_{q_0,\delta}$ for which the conclusion of Theorem 3.2 fails. For example, $(1+\varepsilon)s_{q_0}$ will converge to $s_{q_0}$, and not to $s_{q_1}e^{i\varphi}$. On the other hand, there is a small neighborhood $N$ of $s_{q_0}$ in $\mathcal{U}_{q_0,\delta}$ such that, for any $u^{(0)} \in N \setminus \mathcal{V}_{cs}$, the solution $u_t$ converges to $s_{q_1}e^{i\varphi}$ as $t \to \infty$. Indeed, it is well-known from the theory of dynamical systems that such a trajectory leaves a (bigger) neighborhood of $s_{q_0}$ essentially along the unstable manifold, and thus satisfies the energy condition $H(u_t) < H(s_{q_0})$ for $t$ sufficiently large. In particular, we can formulate the following

**Corollary 3.3.** *The conclusion of Theorem 3.2 holds if $u^{(0)}(x) = s_{q_0}(x) + \varepsilon e^{iq_1 x}$ with $\varepsilon$ sufficiently small.*

**Proof of Theorem 3.2.** Denote by $u_t$ the (unique) solution of Eq.(3.4) in $\mathcal{U}_{q_0,\delta}$ with initial data $u^{(0)}$. As a solution of a parabolic equation with smooth coefficients, $u_t$ is a smooth function of $x, t$ for all $t > 0$. In particular, $H(u_t)$ is a $\mathcal{C}^1$ function of $t$, and an easy calculation shows that

$$\frac{d}{dt}H(u_t) = -\frac{1}{T}\int_0^T \left|u_t'' + u_t - |u_t|^2 u_t\right|^2 dx = -\frac{1}{T}\int_0^T |\dot{u}_t|^2 dx ,$$

where $T = 2\pi/\delta$. Since $H(u) \geq -1/4$ for all $u \in \mathcal{U}_{q_0,\delta}$, we see that $H(u_t)$ is decreasing and bounded from below, hence converges as $t \to \infty$ to some value $H_\infty$.

On the other hand, it is well-known that the trajectory $\{u_t\}_{t\geq 0}$ is precompact in $\mathcal{U}_{q_0,\delta}$, see for example [Te]. In particular, its $\omega$-limit set $\Omega$ is non-empty, and any $u \in \Omega$ is an equilibrium point of the system satisfying $H(u) = H_\infty < H(s_{q_0})$, see [Ha]. By Proposition 3.1, $\Omega$ is thus contained in the circle of fixed points $s_{q_1}e^{i\varphi}$, $\varphi \in [0, 2\pi]$. Now, a straightforward application of the stable manifold theorem shows that this circle has a small attractive neighborhood $\mathcal{A}$ in $\mathcal{U}_{q_0,\delta}$, and that every trajectory of the system entering $\mathcal{A}$ converges to a point of the circle as $t \to \infty$ [EG]. By definition of the $\omega$-limit set, this is the case for the trajectory $u_t$, and the proof of Theorem 3.2 is complete.

**Proof of Proposition 3.1.** We first recall that the stationary solutions $u(x)$ of the Ginzburg-Landau equation form a two-parameter family which can be indexed by the values of two "first integrals" $J$ and $E$, see [DGJ, BR, Ga]. If $u(x)$ is written in polar coordinates $r(x)e^{i\varphi(x)}$, then $E$ and $J$ can be expressed as

$$J = r^2(x)\varphi'(x), \quad E = \tfrac{1}{2}|u'|^2(x) + \tfrac{1}{2}|u|^2(x) - \tfrac{1}{4}|u|^4(x)$$
$$= \tfrac{1}{2}r'(x)^2 + \tfrac{1}{2}\frac{J^2}{r^2(x)} + \tfrac{1}{2}r^2(x) - \tfrac{1}{4}r^4(x) . \tag{3.5}$$

Note that these formulas allow for an interpretation of $E$ and $J$ as the energy and the angular momentum of a point particle (in $\mathbf{R}^2$) moving in the radially symmetric potential $V(r) =$ $\frac{1}{2}r^2 - \frac{1}{4}r^4$. In this mechanical analogy, the role of the time is played by the space variable $x \in \mathbf{R}$.

As is well-known (see, e.g, [Ga]), the only stationary solutions of Eq.(3.4) whose modulus $|u(x)|$ is constant in $x$ are the functions $s_q(x) = \sqrt{1-q^2}\, e^{iqx}$ for $q \in [-1,1]$. Since $\delta > 1$, it is easy to verify that $s_q \in \mathcal{U}_{q_0,\delta}$ if and only if $q = q_0$, $q = q_1$, or $q = 1$. In this last case, $s_q \equiv 0$ and $0 = H(s_q) > -\frac{1}{4}(1-q_0^2)^2 = H(s_{q_0})$. So, we see that Proposition 3.1 holds if we consider only stationary solutions with constant modulus.

To conclude the proof, we need therefore only show that the space $\mathcal{U}_{q_0,\delta}$ contains no stationary solution of Eq.(3.4) whose modulus is periodic in $x$ and non-constant. If $u(x) = r(x)e^{i\varphi(x)}$ is any such solution and if $J, E$ are the corresponding values of the first integrals (3.5), then the (minimal) period of the modulus $r(x)$ is given by

$$T(E, J) = 2 \int_{r_1(E,J)}^{r_2(E,J)} \frac{dr}{\left(2E - J^2 r^{-2} - r^2 + \frac{1}{2}r^4\right)^{1/2}}, \qquad (3.6)$$

where $r_1$ and $r_2$ are the two smallest roots of the denominator, see [DGJ]. In the mechanical analogy above, this period corresponds to the time between two minima (or maxima) of the motion of the particle. Another important quantity is the total increase of the phase $\varphi(x)$ during a period $T$. It is given by

$$\eta(E, J) = 2 \int_{r_1(E,J)}^{r_2(E,J)} \frac{J}{r^2} \frac{dr}{\left(2E - J^2 r^{-2} - r^2 + \frac{1}{2}r^4\right)^{1/2}}, \qquad (3.7)$$

and corresponds to the "advance of the perihelion" in the mechanical analogy.

**Remark.** Strictly speaking, the formulas (3.6), (3.7) make sense only if $J \neq 0$. If $J = 0$, then $u(x)$ is real (up to a global phase factor) and it is easy to verify that such a function cannot belong to $\mathcal{U}_{q_0,\delta}$ if $(q_0, \delta) \in \hat{\mathcal{E}}$, except for the trivial case $u \equiv 0$ which we have already excluded. Thus, we shall always assume that $J \neq 0$.

It follows easily from the definitions that a stationary solution $u(x) = r(x)e^{i\varphi(x)}$ of Eq.(3.4) parametrized by $E, J$ belongs to $\mathcal{U}_{q_0,\delta}$ if and only if

$$jT(E, J) = \frac{2\pi}{\delta} \quad \text{and} \quad j\eta(E, J) = \frac{2\pi q_0}{\delta} \bmod 2\pi, \qquad (3.8)$$

for some $j \in \mathbf{N}$. In other words, the modulus $r(x)$ has to be periodic of (not necessarily minimal) period $2\pi/\delta$, and the increase of the phase $\varphi(x)$ over a period must be equal to $q_0(2\pi/\delta)$ modulo $2\pi$. However, the next two lemmas will show that this is *not* possible if $(q_0, \delta) \in \hat{\mathcal{E}}$.

**Lemma 3.4.** *Let $E, J$ be as above, and let $0 < y_1 < y_2 < y_3$ be the roots of the polynomial*

$$y^3 - 2y^2 + 4Ey - 2J^2 = (y - y_1)(y - y_2)(y - y_3). \qquad (3.9)$$

*Then the period $T(E, J)$ satisfies*

$$T > \frac{2\pi}{\sqrt{3y_3 - 2}}. \qquad (3.10)$$

*In particular, since $y_1 + y_2 + y_3 = 2$, one always has $y_3 < 2$, hence $T > \pi$.*

**Lemma 3.5.** *Let $E, J$ as above, and assume that $\delta = 2\pi/T(E,J) > 1$. Then the phase increase $\eta(E,J)$ satisfies*

$$\pi < |\eta| < \pi\sqrt{\frac{2}{3} + \frac{4}{3\delta^2}} \ . \tag{3.11}$$

*In particular, one always has $\pi < |\eta| < \sqrt{2}\pi$.*

Using these two lemmas (which will be proved in the Appendix) we can now conclude the proof of Proposition 3.1. Let $(q_0, \delta) \in \hat{\mathcal{E}}$, and assume that there exists a stationary solution $u \in \mathcal{U}_{q_0,\delta}$ of the GL equation (3.4) whose modulus $|u|$ is periodic in $x$ (and non-constant). If $E, J$ are the corresponding values of the first integrals (3.5), then Eq.(3.8) is verified for some $j \in \mathbf{N}$. Since $\delta > 1$ and (by Lemma 3.4) $T > \pi$, the first equality in Eq.(3.8) can only be satisfied if $j = 1$. The second equality then implies that $\eta = 2\pi q_0/\delta + 2\pi m$ for some $m \in \mathbf{Z}$, and by Lemma 3.5 we must have $m = -1$ if $\eta < 0$ and $m = 0$ if $\eta > 0$. Combining these facts with the bounds (3.11), we obtain the inequalities

$$\delta - \sqrt{\frac{1}{3} + \frac{\delta^2}{6}} < q_0 < \sqrt{\frac{1}{3} + \frac{\delta^2}{6}} \ .$$

In particular, we find $\delta > \sqrt{6q_0^2 - 2}$, in contradiction with the definition of $\hat{\mathcal{E}}$. This means that Eq.(3.8) cannot be satisfied if $(q_0, \delta) \in \hat{\mathcal{E}}$, and the proof of Proposition 3.1 is complete.

## 4. The Swift-Hohenberg Equation

In this section, we address the question of phase slips in the Swift-Hohenberg equation

$$\partial_t U(x,t) = \left(3\varepsilon^2 - (1+\partial_x^2)^2\right)U(x,t) - U^3(x,t) \ , \tag{4.1}$$

where $U : \mathbf{R} \times \mathbf{R}^+ \to \mathbf{R}$. For small $\varepsilon > 0$, this equation has a family of spatially periodic stationary states of the form

$$S_{\varkappa,\omega}(x) = 2\varkappa \cos(\omega x) + \mathcal{O}(\varkappa^2) \ , \tag{4.2}$$

where $\omega \approx 1$ and $3\varepsilon^2 = (\omega^2 - 1)^2 + 3\varkappa^2$, see [CE1]. These states are known to be unstable if $(\omega^2 - 1)^2 > \varepsilon^2$ and marginally stable otherwise. Numerical calculations show that, when such unstable states are perturbed, they typically relax by undergoing phase slips which change the number of zeros. A natural setup, which is considered in theory and experiments alike, consists in applying space periodic perturbations to these solutions. Then the phase-slips occur at periodically spaced points, changing in this way the density of zeros. In experiments this jump in density is in fact used to determine the limits of stability of the periodic solutions.

The aim of this section is to present a rigorous description of these phenomena, using the results of the preceding sections for the GL equation. The relation between the two problems is the so-called "approximation property," which says that the evolution of the SH equation is accurately described by the GL equation for a long (but finite) interval of time, see [CE2, vH, KSM, Sch]. A precise formulation of this property is as follows: First, fix $T_0 > 0$, $R > 0$, and let $A(\xi, \tau)$ be a solution of the GL equation

$$\partial_\tau A(\xi, \tau) = \partial_\xi^2 A(\xi, \tau) + A(\xi, \tau) - A(\xi, \tau)|A(\xi, \tau)|^2 , \tag{4.3}$$

satisfying $\|A(\cdot, 0)\|_{\mathcal{C}^4} \leq R$. Next, for any $\varepsilon > 0$, define

$$V_{A,\varepsilon}(x, t) = 2\varepsilon \operatorname{Re} \left( A(\tfrac{\sqrt{3}}{2}\varepsilon x, 3\varepsilon^2 t) e^{ix} \right) , \tag{4.4}$$

and let $U_{A,\varepsilon}(x, t)$ be the unique solution of the SH equation in $\mathcal{C}^4(\mathbf{R})$ with initial condition

$$U_{A,\varepsilon}(x, 0) = V_{A,\varepsilon}(x, 0) . \tag{4.5}$$

Under these assumptions, there exist an $\varepsilon_0 > 0$ and a $C < \infty$ such that, for all $\varepsilon \leq \varepsilon_0$ and all $t \in [0, T_0/\varepsilon^2]$, one has

$$\|U_{A,\varepsilon}(\cdot, t) - V_{A,\varepsilon}(\cdot, t)\|_{L^\infty} \leq C\varepsilon^4 t , \tag{4.6}$$

see [KSM]. In particular, the approximation $V_{A,\varepsilon}(\cdot, t)$ is within $\mathcal{O}(\varepsilon^2)$ of the true solution $U_{A,\varepsilon}(\cdot, t)$ when $t \in [0, T_0/\varepsilon^2]$.

**Remark.** In [KSM], the authors consider the more general case where $U_{A,\varepsilon}(x, 0) = V_{A,\varepsilon}(x, 0) + \mathcal{O}(\varepsilon^2)$, and they show that $\|U_{A,\varepsilon}(\cdot, t) - V_{A,\varepsilon}(\cdot, t)\| \leq C\varepsilon^2$ for all $t \in [0, T_0/\varepsilon^2]$. However, in the particular case where the initial data for $U_{A,\varepsilon}$ and $V_{A,\varepsilon}$ coincide, their proof yields the stronger result (4.6).

We first give a local description of a typical phase slip for the SH equation, using the results of Section 2. Let $A(\xi, \tau)$ be a solution of the GL equation (4.3) satisfying the assumptions of Theorem 2.7 for $u(x, t)$. Without loss of generality, we may assume that $A(0, d) = 0$, i.e. the phase slip occurs exactly at $\xi^* = 0$, $\tau^* = d$. Then, according to Eq.(2.3), there exist $a > 0$, $b \in \mathbf{C}$, $\operatorname{Re} b \neq 0$, and $d > 0$ such that

$$A(\xi, \tau) = \tfrac{1}{2} a\xi^2 + ib\xi + a(\tau - d) + \mathcal{O}\left( |\xi|^3 + |\tau - d|^{3/2} \right) , \tag{4.7}$$

for all $|\xi| \leq d^{1/2}$ and all $\tau \in [0, 2d]$. For $\varepsilon > 0$, we define $V_{A,\varepsilon}(x, t)$ by Eq.(4.4) and we let $U_{A,\varepsilon}(x, t)$ be the solution of the SH equation satisfying the initial condition (4.5). If $d \leq 3T_0/2$ and $\varepsilon$ is sufficiently small, then Eq.(4.6) implies that

$$U_{A,\varepsilon}(x, t) = V_{A,\varepsilon}(x, t) + \mathcal{O}(\varepsilon^4 t) , \tag{4.8}$$

for all $|x| \leq 2(d/3)^{1/2}\varepsilon^{-1}$ and all $t \in [0, 2d/(3\varepsilon^2)]$.

To show that $U(x, t)$ undergoes a phase slip near $x = 0$, $t = d/(3\varepsilon^2)$, we assume from now on that $\varepsilon \leq d/K$ for some (large) constant $K > 0$, and we restrict Eq.(4.8) to the smaller region

$$R_{d,\varepsilon} = \left\{ (x, t) \mid |x| \leq K , |t - d/(3\varepsilon^2)| \leq K/\varepsilon \right\} .$$

Setting $b' \equiv b'_1 + ib'_2 = \sqrt{3}b/2$, we find

$$\begin{aligned}U_{A,\varepsilon}(x,t) &= 2\varepsilon^2 \operatorname{Re}\left((ib'x + a(3\varepsilon t - d/\varepsilon))e^{ix}\right) + \mathcal{O}(\varepsilon^2 d) \\ &= 2\varepsilon^2\left((a(3\varepsilon t - d/\varepsilon) - b'_2 x)\cos x - b'_1 x \sin x\right) + \mathcal{O}(\varepsilon^2 d),\end{aligned} \quad (4.9)$$

uniformly in $R_{d,\varepsilon}$.

Since we wish to study how the number of nodes of the solution changes with time, we look for points where $U_{A,\varepsilon}(x,t) = 0$. The leading term in Eq.(4.9) vanishes if and only if $f_b(x) = a(3\varepsilon t - d/\varepsilon)$, where

$$f_b(x) = b'_1 x \tan x + b'_2 x . \quad (4.10)$$

The graph of the function $f_b(x) - s$ is represented in Fig. 5 for $b'_1 < 0$. We see that if $-s \gg 1$, the roots of the equation $f_b(x) = s$ are located near the points $x = (n + 1/2)\pi$, $n \in \mathbf{Z}$. As $s$ goes through zero, the two central roots coalesce and disappear in a neighborhood of the origin. Conversely, if $b'_1 > 0$, a pair of roots is created near $x = 0$. These phenomena are precisely what we call *phase slips* for the SH equation. They will occur if we choose the constant $K = K(a,b) > 0$ so that the quantity $s = a(3\varepsilon t - d/\varepsilon)$ varies over a sufficiently large range as $t$ varies in the region $R_{d,\varepsilon}$. Of course, when considering the full Eq.(4.9), the picture will be slightly distorted by the error terms, but qualitatively it remains the same if $d$ is sufficiently small. These arguments provide a sketch of the proof of the following

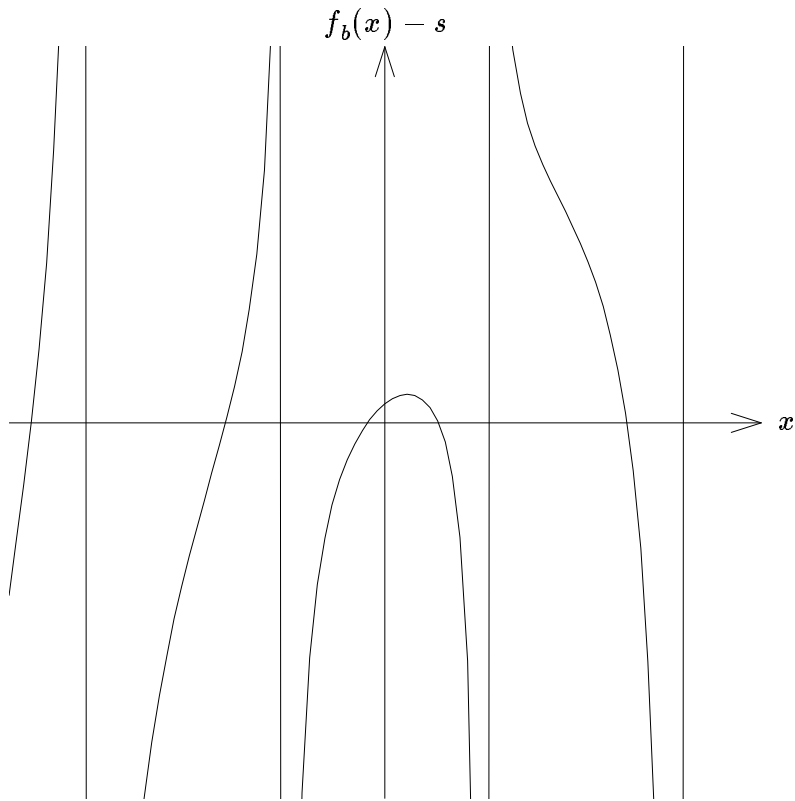

**Fig. 5**: The amplitude of the SH equation, divided by $\cos x$. We show the graph of the function $f_b(x) - s = b'_1 x \tan x + b'_2 x - s$ for $s < 0$.

**Theorem 4.1.** *Assume that $A(\xi,\tau)$ satisfies the assumptions of Theorem 2.7 for $u(x,t)$. For $\varepsilon > 0$, let $U_{A,\varepsilon}(x,t)$ be the solution of the SH equation with initial condition*

$$2\varepsilon \operatorname{Re}\left(A(3^{1/2}\varepsilon x/2, 0)e^{ix}\right) .$$

*Then there exist a $d_1 \leq d^*$ and a $K = K(a,b) > 0$ such that, for all $d \in [0,d_1]$ and all $\varepsilon \leq d/K$, the function $U_{A,\varepsilon}(x,t)$ undergoes a phase slip near $x = 0$, $t = d/(3\varepsilon^2)$. This phase slip corresponds to the creation $(\operatorname{Re} b > 0)$ or the annihilation $(\operatorname{Re} b < 0)$ of a pair of roots of $U_{A,\varepsilon}(x,t)$.*

We next address the question of the global evolution of an Eckhaus instability for the Swift-Hohenberg equation. Assume that $q_0, q_1$ are two wave-numbers such that $(q_0, q_0 - q_1)$ belongs to the region $\hat{\mathcal{E}}$ defined in Section 3, and let $s_j(\xi) = (1 - q_j^2)^{1/2} e^{iq_j \xi}$ for $j = 0, 1$. If $\varepsilon > 0$ is sufficiently small, the SH equation (4.1) has two families of periodic stationary solutions $S_{j,\varphi}(x)$ which are of the form

$$S_{j,\varphi}(x) = 2\varepsilon \operatorname{Re}\left(s_j(\tfrac{\sqrt{3}}{2}\varepsilon x)e^{i(x+\varphi)}\right) + \mathcal{O}(\varepsilon^2) . \tag{4.11}$$

We recall that $S_{0,\varphi}(x)$ is always linearly unstable, whereas $S_{1,\varphi}(x)$ is stable or unstable according to whether $q_1^2 < 1/3$ or $q_1^2 > 1/3$. Combining Theorem 3.2 with the approximation property (4.6), we shall show the following

**Theorem 4.2.** *Assume that $(q_0, q_0 - q_1) \in \hat{\mathcal{E}}$. For any $\rho > 0$, there exist $\varepsilon_0 > 0$, $T_0 > 0$ such that, for all $\varepsilon \in [0,\varepsilon_0]$, the Swift-Hohenberg equation (4.1) has a solution $U(x,t)$ satisfying*

$$\|U(\cdot,0) - S_{0,0}(\cdot)\|_{L^\infty} \leq \varepsilon\rho , \quad \|U(\cdot,T_0/\varepsilon^2) - S_{1,\varphi}(\cdot)\|_{L^\infty} \leq \varepsilon\rho , \tag{4.12}$$

*for some $\varphi \in [0,2\pi)$.*

**Remark.** The construction of $U(x,t)$ will show that $U(x,0)$ and $S_{0,0}(x)$ have the same density of zeros (if $\rho$ is sufficiently small), and similarly for $U(x,T_0/\varepsilon^2)$ and $S_{1,\varphi}(x)$. Since $q_0 \neq q_1$, this means that $U(x,t)$ undergoes phase slips as $t$ varies from 0 to $T_0/\varepsilon^2$.

**Remark.** Since the solutions depend continuously on time, there is in fact a neighborhood of initial data near $U(\cdot,0)$ for which the conclusions of Theorem 4.2 hold. In this sense, the Eckhaus instability leads generically to phase slips.

**Proof.** Fix $\rho > 0$. According to Theorem 3.2, there exists a solution $A(\xi,\tau)$ of the GL equation (4.3) such that $A(\xi,-\infty) = s_0(\xi)$ and $A(\xi,\infty) = e^{i\varphi}s_1(\xi)$, for some $\varphi \in [0,2\pi)$. Therefore, there is a $T_0 > 0$ such that, after choosing the origin of $\tau$ conveniently, one has

$$\|A(\cdot,0) - s_0(\cdot)\|_{L^\infty} \leq \rho/3 , \quad \|A(\cdot,3T_0) - e^{i\varphi}s_1(\cdot)\|_{L^\infty} \leq \rho/3 . \tag{4.13}$$

Furthermore, by the approximation property, there is an $\varepsilon_0 > 0$ such that, if $V(x,t) \equiv V_{A,\varepsilon}(x,t)$ is defined by Eq.(4.4) and if $U(x,t) \equiv U_{A,\varepsilon}(x,t)$ is the solution of the SH equation with initial condition (4.5), then Eq.(4.6) holds for all $\varepsilon \leq \varepsilon_0$. In view of Eqs.(4.6), (4.11), (4.13), we thus have

$$\begin{aligned}\|U(\cdot,0) - S_{0,0}(\cdot)\|_{L^\infty} &= \|V(\cdot,0) - S_{0,0}(\cdot)\|_{L^\infty} \\ &\leq 2\varepsilon \|A(\cdot,0) - s_0(\cdot)\|_{L^\infty} + \mathcal{O}(\varepsilon^2) \leq 2\varepsilon\rho/3 + C^*\varepsilon^2 \ ,\end{aligned}$$

and similarly

$$\begin{aligned}\|U(\cdot,T_0/\varepsilon^2) - S_{1,\varphi}(\cdot)\|_{L^\infty} &\leq \|U(\cdot,T_0/\varepsilon^2) - V(\cdot,T_0/\varepsilon^2)\|_{L^\infty} + \|V(\cdot,T_0/\varepsilon^2) - S_{1,\varphi}(\cdot)\|_{L^\infty} \\ &\leq C\varepsilon^2 + 2\varepsilon\|A(\cdot,3T_0) - e^{i\varphi}s_1(\cdot)\|_{L^\infty} + \mathcal{O}(\varepsilon^2) \\ &\leq C^*\varepsilon^2 + 2\varepsilon\rho/3 \ .\end{aligned}$$

Therefore, if $C^*\varepsilon < \rho/3$, the solution $U(x,t)$ verifies Eq.(4.12). The proof of Theorem 4.2 is complete.

**Remark.** Unfortunately, it is not possible to conclude from the preceding arguments that the solution $U(x,t)$ actually converges to $S_{1,\varphi}(x)$ as $t \to \infty$. Indeed, the approximation (4.6) is only valid over a finite interval of time, whereas the trajectory $A(\xi,\tau)$ takes an infinite time to reach the fixed point $e^{i\varphi}s_1(\xi)$. Moreover, even when $q_1^2 < 1/3$, we do not know if the stationary solution $S_{1,\varphi}(x)$ of the SH equation is non-linearly stable.

# Appendix

**Proof of Lemma 3.4.** We first rewrite the expression for the period $T$ in a more convenient form. Setting $y = r^2$ in Eq.(3.6), we find

$$T = \int_{y_1}^{y_2} \frac{dy}{\sqrt{2Ey - J^2 - y^2 + y^3/2}} = \sqrt{2} \int_{y_1}^{y_2} \frac{dy}{\sqrt{(y-y_1)(y_2-y)(y_3-y)}} \ .$$

Using the change of variables

$$y = a(\psi) \equiv y_1 \cos^2(\psi) + y_2 \sin^2(\psi) \ , \quad dy = 2\sqrt{(y-y_1)(y_2-y)}\, d\psi \ ,$$

we obtain

$$T = 2\sqrt{2} \int_0^{\pi/2} \frac{d\psi}{\sqrt{y_3 - a(\psi)}} = \frac{\pi\sqrt{2}}{\mathcal{M}(\sqrt{y_3-y_1}, \sqrt{y_3-y_2})} \ ,$$

where $\mathcal{M}(a,b)$ is the arithmetic-geometric mean of $a$ and $b$, see [BB]. By the definition of this mean, we have $\mathcal{M}(a,b) \leq (a+b)/2$, so that

$$T \geq \frac{2\pi\sqrt{2}}{\sqrt{y_3-y_1} + \sqrt{y_3-y_2}} \ . \tag{A.1}$$

It remains to show that Eq.(A.1) implies Eq.(3.10). The idea is to fix $y_3$ and to consider the right-hand side of Eq.(A.1) as a function of $y_1$ alone, using the fact that $y_2 = 2 - y_1 - y_3$. Thus, we define

$$F(y_1) = \sqrt{y_3 - y_1} + \sqrt{y_3 - y_2} = \sqrt{y_3 - y_1} + \sqrt{2y_3 - 2 + y_1},$$

and an easy calculation shows that $F'(y_1) > 0$. Now, since $y_1 < y_2$ and $y_1 + y_2 + y_3 = 2$, we always have $y_1 < 1 - y_3/2$, hence $F(y_1) < F(1 - y_3/2) = \sqrt{6y_3 - 4}$. Inserting this bound in Eq.(A.1), we find

$$T > \frac{2\pi\sqrt{2}}{\sqrt{6y_3 - 4}} = \frac{2\pi}{\sqrt{3y_3 - 2}},$$

as asserted. The proof of Lemma 3.4 is complete.

**Proof of Lemma 3.5.** Since the GL equation (3.4) is invariant under the complex conjugation $u \to \bar{u}$, there is no loss of generality in assuming that $J > 0$, hence $\eta > 0$. Proceeding as in the proof of Lemma 3.4, we rewrite the expression (3.7) of $\eta$ as

$$\eta = 2\sqrt{2} \int_0^{\pi/2} \frac{J}{a(\psi)} \frac{d\psi}{\sqrt{y_3 - a(\psi)}}, \tag{A.2}$$

where again $a(\psi) = y_1 \cos^2(\psi) + y_2 \sin^2(\psi)$.

To prove the lower bound on $\eta$, we first note that $y_1 y_2 y_3 = 2J^2$, in view of Eq.(3.9). On the other hand, an elementary calculation gives

$$\frac{2}{\pi} \int_0^{\pi/2} \frac{d\psi}{a(\psi)} = \frac{1}{\sqrt{y_1 y_2}}. \tag{A.3}$$

Using Eq.(A.2), we obtain

$$\eta > 2\sqrt{y_1 y_2 y_3} \int_0^{\pi/2} \frac{d\psi}{a(\psi)\sqrt{y_3}} = \pi.$$

To prove the upper bound on $\eta$, we apply the Schwarz inequality to the right-hand side of Eq.(A.2) and obtain the bound

$$\eta \leq \pi\sqrt{2} J \left( \frac{2}{\pi} \int_0^{\pi/2} \frac{d\psi}{a(\psi)} \right)^{1/2} \left( \frac{2}{\pi} \int_0^{\pi/2} \frac{d\psi}{a(\psi)(y_3 - a(\psi))} \right)^{1/2}.$$

Using Eq.(A.3) and the analogous result

$$\frac{2}{\pi} \int_0^{\pi/2} \frac{d\psi}{a(\psi)(y_3 - a(\psi))} = \frac{1}{y_3} \left( \frac{1}{\sqrt{y_1 y_2}} + \frac{1}{\sqrt{y_3 - y_1}\sqrt{y_3 - y_2}} \right),$$

we thus find

$$\eta \leq \pi \left(1 + \sqrt{\frac{y_1 y_2}{(y_3 - y_1)(y_3 - y_2)}}\right)^{1/2} . \qquad (A.4)$$

It remains to show that Eq.(A.4) implies the upper bound in Eq.(3.11). First, we note that Eq.(3.10) implies $\sqrt{3y_3 - 2} > 2\pi/T = \delta$, hence $y_3 > (2 + \delta^2)/3$. Since $\delta > 1$ by assumption, we see that $y_3 > 1$ too. Next, proceeding as above, we fix $y_3 \in (1, 2)$ and define

$$G(y_1) = \frac{y_1 y_2}{(y_3 - y_1)(y_3 - y_2)} = \frac{y_1(2 - y_1 - y_3)}{(y_3 - y_1)(2y_3 - 2 + y_1)} .$$

By direct calculation, we find

$$G'(y_1) = \frac{2(y_2 - y_1)y_3(y_3 - 1)}{(y_3 - y_1)^2(y_3 - y_2)^2} > 0 ,$$

so that $G(y_1) < G(1 - y_3/2) = (2 - y_3)^2/(3y_3 - 2)^2$. Since this expression is a decreasing function of $y_3$, we can replace $y_3$ by the lower bound $(2 + \delta^2)/3$. We thus obtain

$$G(y_1) = \frac{y_1 y_2}{(y_3 - y_1)(y_3 - y_2)} < \left(\frac{4 - \delta^2}{3\delta^2}\right)^2 . \qquad (A.5)$$

Combining Eqs.(A.4), (A.5), we find

$$\eta < \pi\sqrt{1 + \frac{4 - \delta^2}{3\delta^2}} = \pi\sqrt{\frac{2}{3} + \frac{4}{3\delta^2}} ,$$

as asserted. The proof of Lemma 3.5 and hence of the global picture for the GL equation are complete.

**Acknowledgments.** This work was started in the stimulating atmosphere of the Mittag-Leffler Institute. We thank it, the Fonds National Suisse, and the NSF Grant DMS-9203359 for financial support.